\documentclass[12pt]{article}
\usepackage{graphicx}
\usepackage{amssymb,amsmath,amsfonts,palatino,amsthm}
\usepackage{amssymb}
\usepackage{epstopdf}
\newcommand{\f}{\begin{equation}}
\newcommand{\ff}{\end{equation}}

\begin{document}

\title{Linking shape dynamics and loop quantum gravity\\}
\author{Lee Smolin\thanks{lsmolin@perimeterinstitute.ca} 
\\
\\
Perimeter Institute for Theoretical Physics,\\
31 Caroline Street North, Waterloo, Ontario N2J 2Y5, Canada}
\date{\today}
\maketitle

\begin{abstract}
Shape dynamics is a reformulation of general relativity, locally equivalent to Einstein's theory, in which the refoliation invariance of the older theory is traded for local scale invariance.  Shape dynamics is here derived in a formulation related to the Ashtekar variables by beginning with a modification of the Plebanski action.  The constraints of shape dynamics and their algebra are reproduced in terms of these new variables.

Revised May 20, 2017.

\end{abstract}

\tableofcontents

\newpage

\section{Introduction}

Many of the challenges of quantum gravity can be tied to the invariance of general relativity under space-time diffeomorphisms.  The aspect of this gauge symmetry which causes difficulties is the consequent freedom to choose an arbitrary time coordinate which slices the space-time manifold into a one parameter family of spatial slices.  Called many fingered time, or refoliation invariance, this gauge symmetry has a different relation to dynamics than other better understood gauge symmetries in that it is generated by a constraint-the Hamiltonian constraint, $\cal H$, which is quadratic in canonical momenta of the gravitational field. This means it does not correspond to a vector field on the configuration space, as is the case with Yang-Mills gauge transformations, as well as the constraints which generate diffeomorphisms of the spatial slices.  Indeed, the freedom to refoliate is locally indistinguishable from the dynamical evolution of local fields, so this gauge symmetry is intertwined with dynamics.  In a word, kinematics is not distinguishable from dynamics, except when a gauge has been fixed.  
This leads to the bundle of technical and conceptual challenges known as the problem of time in quantum gravity.

One aspect of this mystery is the holographic nature of general relativity according to which the Hamiltonian of general relativity is a boundary term when acting on solutions to the constraints\cite{ADM}.  This makes key properties such as the positivity of energy and the stability of the ground state highly non-trivial to demonstrate, classically\cite{SY,Witten} as well as 
quantum mechanically\cite{ls-artem,ls-positive}.

At the classical level, while posing challenges to our understanding,  this intertwining of dynamics and kinematics is clearly correct, and has been largely understood.  But it can be questioned whether fundamentally quantum theory can be made sense of in the absence of a preferred time variable.  After all, there are physical predictions of linearized quantum general relativity, in which we quantize the linearized modes of the theory on a fixed background such as Minkowski or DeSitter spacetime, that depend on a preferred class of time coordinates.  These include effects of operator ordering, which depend on a splitting of modes into positive and negative frequency-positive and negative with respect to a preferred class of clocks.  As shown by \cite{parityb} these include the possibility of parity breaking in the production of tensor modes in inflation.   How are these physical effects to emerge from the low energy limit of a fully non-perturbative background independent quantum theory of gravity if the latter does not also depend on a preferred class of time variables?

One attractive option is then to posit that the refoliation invariance of general relativity is only an approximate or effective gauge invariance, applicable only in the low energy or classical limit of a quantum theory that is defined with respect to  a preferred slicing-or class of slicings-of space-time.  The possible advantages of this have been shown from several points of view 
including Horava-Lifshitz theory\cite{HL} and causal dynamical triangulations\cite{CDT}.  One advantage is that it allows parity to be broken in the production of tensor modes in early universe cosmology\cite{parityb}.  For many reasons, a theory with a preferred time coordinate and a bulk Hamiltonian fits better into the structure of quantum theory\footnote{For the case for time playing a preferred role in fundamental physics, see \cite{timebooks,TN,time-pt}.}.  

However, it does not seem correct to merely drop the Hamiltonian constraint, as that introduces a new, scalar degree of freedom, whose influence must be suppressed to reproduce the predictions of general relativity. For it is the Hamiltonian constraint that is responsible for gravitational waves being pure spin two.   Moreover the Newtonian limit of the Hamiltonian constraint contains the Poisson equation for the gravitational potential, $\phi$, 
\f
\nabla^2 \phi = 4 \pi G \rho,
\label{Poisson}
\ff 
and, without that, a theory may be relativistic, but it is not a theory of gravity!  

Consequently, an even more attractive option is to  trade the refoliation invariance for another local gauge invariance in such a way that the resulting theory has the same physical degrees of freedom as general relativity and, in spite of loosing the invariance under many fingered time, is locally equivalent to general relativity.  This idea is realized in a theory  called {\it shape dynamics\cite{SD1,SD2}.}  

Shape dynamics has been constructed and understood starting with the canonical formulation of general relativity by following three steps\cite{linking}.  

We begin with the phase space of the $ADM$ formulation of Hamiltonian general relativity\cite{ADM} on 
a compact surface, $\Sigma$, whose degrees of freedom
are given by the three metric $q_{ab}$ and canonical momenta, $\tilde{\pi}^{ab}$, in terms of the  Hamiltonian and diffeomorphsm constraints, ${\cal H}$ and ${\cal D}_a$.

\begin{enumerate}

\item{}Enlarge the phase space of the theory by adding a scalar degree  
degree of freedom, $\psi$ and its canonical momenta $\tilde{\pi}_\psi$.  

\item{}Add a new first class constraint, $\cal S$ that expresses the fact that $\psi$  and its momenta
are redundant variables.

The theory with enlarged phase space, $\Gamma^{ext}=\{q_{ab},\psi;  \tilde{\pi}^{ab}, \tilde{\pi}_\psi  \}$ and system 
of constraints, ${\cal C} = ({\cal H}, {\cal S}, {\cal D}_a )$ is called {\it the linking theory\cite{linking}.}  The total system of constraints is first class.


\item{}One shows that the theory defined by the system $(\Gamma^{ext}, {\cal C})$ can be gauge fixed two ways.  One can 
gauge fix $\cal S$ by imposing $\psi =0$ which reduces the theory to general relativity.  Or one can gauge fix
$\cal H$ by imposing $\tilde{\pi}_\psi = c \sqrt{det (q)}$, where $c$ is a possibly slice dependent constant, which 
breaks refoliation invariance and leads to an equivalent formulation of the theory with a new gauge invariance generated 
by $\cal S$.  

\end{enumerate}

There is then a central theorem which holds that there is only a single way to construct a linking theory based on the extended  $ADM$ phase space, invariant under spatial diffeomorphism (and subject to some technical assumptions)\cite{HG-unique}. That unique theory is given by choosing the pair 
$({\cal H}, {\cal S})$, 
where $\cal H$ is the ADM Hamiltonian constraint and 
\f
{\cal S}= \tilde{\pi}_\psi - q_{ab} \tilde{\pi}^{ab}
\ff
is the generator of local scale transformations.  Moreover,  when the gauge fixing $\tilde{\pi}_\psi =c \sqrt{det (q)}$ is chosen,
\f
{\cal S}_{\tilde{\pi}_\psi =c \sqrt{det (q)}}= c \sqrt{det (q)} - q_{ab} \tilde{\pi}^{ab} =0
\ff
which is the constant-mean-curvature gauge condition (CMC slicing).  

Shape dynamics has already illuminated longstanding issues in gravitational physics such as the origin of irreversibility in the universe\cite{SD-time} and the reasons for the $AdS/CFT$ correspondence\cite{SD-AdS}.  And, prompted by some promising observations\cite{Tim-LSD}, the quantum theory is now the focus of current work.  

In this contribution we develop shape dynamics by addressing two questions which may open the way for progress on its quantization.

\begin{itemize}

\item{}We derive shape dynamics for the more modern formulations of general relativity where the configuration space is a space of connections rather than metrics\footnote{Other approaches to expressing shape dynamics in a connection formulation are in\cite{others}.}.  This allows general relativity to be studied and quantized with methods derived from quantum gauge theories such as loop quantum gravity and spin foam models.  We study here the maximally chiral Ashtekar variables\cite{AA}, although the results may work for arbitrary connection based theories.

\item{}We derive shape dynamics beginning with an action principle and extending the configuration space and lagrangian, as opposed to starting with an extension of the Hamiltonian theory.  In particular we make use of Plebanski's action principle\cite{Plebanski,CDJ}, which being cubic in local fields, is the simplest possible form in which the dynamics of general relativity may be expressed.  

\end{itemize}

Both these results should open paths to the quantization of shape dynamics,  by making contact with modern techniques in quantum gravity that depend on both connection variables and action principles.  Some very tentative observations about this are in the conclusions.  Before that, in the next section we construct the linking theory for shape dynamics expressed in the Ashtekar variables by starting with an extension of the Plebanski action and deriving the extended phase space and constraint algebra.  In section 3 we see how this formulation of shape dynamics arises by gauge fixing the linking theory.

\section{The linking theory}

We begin with the chiral Plebanski action\cite{Plebanski,CDJ},
\f
S= \int_{\cal M} B^i \wedge F_i -\frac{1}{2} \phi_{ij} B^i \wedge B^j   ,
\label{plebanski}
\ff
where $B^i$ is an $SU(2)$ valued two form,  $F^i= dA^i +\frac{1}{2} \epsilon^{ijk} A_j \wedge A_k $ is the curvature  of an $SU(2)$ connection, $A^i$ and the symmetric matrix of scalar fields
$\phi^{ij} $ is restricted by the condition
\f
\phi_{ii} = -3 \Lambda
\ff
We will here consider the case that all fields are real, which means that we are describing general relativity with Euclidean signature.  The case of Lorentzian gravity involves some subtleties which will be discussed elsewhere.   

To construct the linking theory we add a scalar field $\psi$ and rescale
\f
B^i \rightarrow e^\psi B^i,  \ \ \ \ \ \   A^i \rightarrow e^{-\psi} A^i,   \ \ \ \ \ \ \phi_{ij} \rightarrow \phi_{ij}
\label{rescale1}
\ff
The action is now,
\f
S= \int_{\cal M} B^i \wedge f_i - \frac{e^{2\psi}}{2} \phi_{ij} B^i \wedge B^j 
\label{plebanski2}
\ff
where the rescaled curvature is
\f
f^i = dA^i +\frac{e^{-\psi}}{2} \epsilon^{ijk} A_j \wedge A_k -A^i \wedge d \psi
\ff

\subsection{Equations of motion}

We vary the fields in (\ref{plebanski2}) to find the equations of motion of the inking theory:
\begin{eqnarray}
\frac{\delta S }{\delta B^i} :  &&  f^i =  e^{ 2\psi} \phi^{ij} B_j
\\
\frac{\delta S }{\delta A^i} :  & &  {\cal D} \wedge B^i = dB^i - e^{-\psi} \epsilon^{ijk} A_j \wedge B_k - B^i \wedge d \psi =0 
\\
\frac{\delta S }{\delta \phi^{ij}} :  & & B^i \wedge B^j - \frac{1}{3} \delta^{ij} B^k \wedge B_k =0 
\\
\frac{\delta S }{\delta \psi } :  & &  e^{2 \psi} = - \frac{ d ( B^i \wedge A_i ) -  \frac{1}{2} e^{-\psi}  \epsilon_{ijk} B^i \wedge A^j \wedge A^k }{\phi_{ij} B^i \wedge B^j }
\label{eom}
\end{eqnarray}

\subsection{The canonical theory}

We proceed to construct the canonical theory.  The first step  is to define the canonical momenta.  As in the usual theory we have
\f
\pi^{ai} = \epsilon^{abc} B_{bc}^i  , \  \  \  \  \  \  \pi_\phi =0
\ff
There is a new canonical momentum for $\psi$:
\f
\pi_\psi = - \pi^a_i A_a^i .
\ff
This gives rise to a new constraint
\f
{\cal S}= \pi_\psi + \pi^a_i A_a^i  =0 . 
\ff
Note that ${\cal S (\rho )} = \int_\Sigma {\cal S} \rho$ generates the action of Weyl transformations, under
\f
\delta_\rho \Phi = \{ \Phi , {\cal S}(\rho ) \}
\ff
we have
\f
\delta_\rho A_a^i =  \rho A_a^i , \ \ \ \  \delta_\rho \pi^a_i  =  - \rho \pi^a_i , \ \ \ \  \delta_\rho \psi = \rho , \ \ \ \ \delta_\rho \pi_\psi =0 . 
\ff
We then rewrite the action in the totally constrained Hamiltonian form
\f
S= \int dt \int_\Sigma  \left ( \pi^a_i \dot{A}_a^i + \pi_\psi \dot{\psi} - A^i_0 {\cal G}^i  - B_{0a}^i {\cal J}^a_i - \rho {\cal S}
\right )
\ff
where the Gauss's law constraint is,
\f
{\cal G}^i = - \partial_a \pi^a_i + e^{-\psi} \epsilon^{ijk} A_{a j } \pi^a_k -  \pi^{a i} \partial_a \psi =0
 \ff
and there are several second class constraints
\f
{\cal J}^a_i = \epsilon^{abc} \left (   f_{bc}^i - e^{2 \psi } \phi^{ij} B_{bc j }
\right ) =0 
\ff

These imply four first class constraints, which follow from the symmetry and trace fixed properties of the $\phi^{ij}$.  These are
easily seen to be the obvious modifications of the Hamiltonian
\f
{\cal H} = \epsilon^{ijk} \left (   \pi^a_i \pi^b_j f_{ab k} - 3 \Lambda e^{2 \psi} \epsilon_{abc} \pi^a_i \pi^b_j \pi^c_k 
\right ) =0
\ff
and  vector constraints
\f
{ \cal V}_a = \pi^b_i f_{ab}^i 
\ff
We can combine ${\cal V}_a$ with the Gauss law constraints to make the spatial diffeomorphism constraints
\f
{\cal D}_a = {\cal V}_a -A_a^i {\cal G}_i 
\ff
This generates spatial diffeomorphisms as
\f
{\cal D} (v) =\int_\Sigma v^a {\cal D}_a =  \int_\Sigma \left (  \pi^a_i {\cal L}_v A_a^i + \pi_\psi {\cal L}_v \psi 
\right ) 
\ff

 \subsection{The shifted Gauss's law and shifted Wilson loops}

Let us take a moment to understand the Gauss's law constraint.  It generates the modified gauge transformation,
\f
\delta_\lambda A_a^i = \{ A^a_i , {\cal G}(\rho ) \} = -\partial_a \lambda^i + e^{-\psi} \epsilon^{ijk} A_{a j} \lambda_k + \lambda_i \partial_a \psi
\ff
To understand this let us undo the transformation (\ref{rescale1}) and write
\f
{\cal A}_a^i = e^{-\psi} A_a^i.
\ff
Let us consider ${\cal A}_a^i$ a composite field which is a function of $A_a^i$ and $\psi$.  ${\cal A}_a^i$ transforms like
a gauge field under 
\f
\delta_{e^{-\psi} \lambda} {\cal A}_a^i = - \partial_a (e^{-\psi} \lambda^i )+ \epsilon^{ijk} {\cal A}_{a j} ( e^{-\psi} \lambda_k ) . 
\ff
It makes sense to then define the rescaled Gauss's law constraint
\f
\tilde {\cal G}^i = e^{-\psi } {\cal G}^i  = e^{-\psi }  \left (   - \partial_a \pi^a_i + e^{-\psi} \epsilon^{ijk} A_{a j } \pi^a_k + \pi^{a i} \partial_a \psi \right ) = 0
 \ff
In other words we consider the shifted gauge transformations,
\f
\tilde{\delta}_\lambda A_a^i = \{ A^a_i , \tilde{\cal G}(\rho ) \}  = \{ A^a_i , {\cal G}(e^{-\psi} \rho ) \} 
\ff

We can then define a gauge invariant Wilson loop observable labeled by a loop $\gamma$ in $\Sigma$
\f
T[ \gamma ; A, \psi ] = Tr [ P e^{\int_\gamma  e^{-\psi } A } ] = Tr [ P e^{\int ds   {\cal A}_a^i \tau_i \dot{\gamma}^a (s) ds }  ] 
\ff

We can also define a shifted flux operator, which depends on a surface, $S$, on which there is a 
Lie algebra valued function, $v^i$,
\f
{\cal E}[S, v^i ] = \int_S d^2 \sigma_a v^i (\sigma ) e^\psi \tilde{E}^a_i (\sigma )
\ff

$T[ \gamma ; A, \psi ]$ and ${\cal E}[S, v^i ]$ commute with ${\cal S}$ and obey the ordinary holonomy-flux algebra.  This is not surprising as they are the original, unshifted fields.

We see that $e^\psi$ can in some respects be thought of as a spacetime dependent Barbero-Immirzi parameter.

\subsection{The algebra of constraints}

We check the algebra of constraints.  The first non-trivial check to make is that the Poisson bracket of the Weyl constraint $\cal S$ 
with the Gauss's law constraint is first class:
\f
\{ {\cal G} (\lambda^i ) , {\cal S}(\rho)  \} = {\cal G} (\rho \lambda^i )  
\ff
This is nontrivial because $\cal S$ has a naked $A_a^i$ that one might presume breaks gauge invariance.

We can also check that the modified Gauss's law constraint still is first class, but the $SU(2)$ structure constants,
$\epsilon^{ijk}$ have become structure functions $e^{-\psi} \epsilon^{ijk}$, so that

\f
\{ {\cal G} (\lambda^i ) , {\cal G}(\mu^i )  \} = {\cal G} ( e^{-\psi} \epsilon^{ijk} \lambda_j \mu_k  )  
\ff

One can also verify that $\cal S$ with ${\cal D}(v)$ is first class:
\f
\{ {\cal S}(\rho), {\cal D} (v)   \} = {\cal S}({\cal L}_v \rho)  
\ff
and that $\cal S$ Poisson commutes with itself 
\f
\{ {\cal S}(\rho), {\cal S}(\sigma )   \} = 0 .
\ff

In addition, , $\cal S$ and $\cal H$  weakly commute, 
\f
\{ {\cal H}( N), {\cal S}(\rho )   \} = {\cal H}(N \rho ).
\ff

Thus, the eight constraints,  ${\cal G}^i , {\cal D}_a $, ${\cal H}$ and $\cal S$ form a first class system.

To summarize, we have defined the linking theory defined by the extend phase space,
\f
\Gamma^{ext } = (A^i_a, \psi ; \tilde{\pi}^a_i, \tilde{\pi} _\psi ),
\ff
on which there is defined the system of eight first class constraints  per point of $\Sigma$
\f
{\cal C} = ({\cal H}, {\cal S}, {\cal D}_a, {\cal G}^i).
\ff

\section{Back to GR or SD by gauge fixing the linking theory}

We may now proceed  to gauge fix either $\cal S$ or $\cal H$.  This gets us either to general relativity 
or shape dynamics.

One can gauge fix $\cal S$ by the condition $\psi =0$, this returns the theory to general relativity.  $\cal S$ is then trivially
solved to set $\pi_\psi $ equal to $- \pi_i^a A_a^i $.

The other alternative is to gauge fix $\cal H$ by imposing the condition 
\f
\pi_\psi =0
\ff 
In this case $\cal S$ becomes
\f
{\cal S} = \tilde{\pi} = \pi^a_i A_a^i =0
\label{max}
\ff
This is analogous to maximal slicing.

Alternatively we can choose the analogy to $CMC$ slicing by choosing to gauge fix the Hamiltonian constraint by imposing, 
\f
\xi = \pi_\psi + <\tilde{\pi} >   \sqrt{ det( \tilde{\pi}^a_i ) }=0  ,
\label{CMCgauge}
\ff 
where 
\f
\tilde{\pi} = \pi^a_i A_a^i  .
\ff 
and
\f
<\tilde{\pi} > = \frac{\int \tilde{\pi} }{V}.
\ff
where $V= \int \sqrt{ det( \tilde{\pi}^a_i ) }$.

In the presence of (\ref{CMCgauge}) the constraint $\cal S$ becomes
\f
\tilde{\pi} =   <\tilde{\pi} >   \sqrt{ det( \tilde{\pi}^a_i ) } . 
\ff
which implies the mean curvature, $p= \frac{\tilde{\pi}}{\sqrt{ det( \tilde{\pi}^a_i ) }}$ is a constant on each
spatial hypersurface.

In this case $\cal H$ must be solved to express $\psi$ as a function of $A_a^i$ and $\pi^a_i$.
This yields a first order partial differential equation to integrate to find $\psi$ 
(at $\Lambda =0$)
\f
w^a \partial_a e^\psi  = e^{-\psi} (\pi^a_i A_a^j \pi^b_j A_b^i - \tilde{\pi}^2 ) + \epsilon^{ijk} \pi^a_i \pi^b_j \partial_a A_{b k} 
\label{weq}
\ff
where the vector field $w^a$ is
\f
w^a = \epsilon^{ijk} \pi^a_i \pi^b_j  A_{b k} 
\ff

To complete the gauge fixing the lapse $N$ is fixed by the condition  that
the gauge fixing condition (\ref{CMCgauge}) is preserved by the Hamiltonian
\f
H = \int_\Sigma N {\cal H} 
\ff
$0= \{ \xi , H \} $ results in a complicated equation for $N$
\begin{eqnarray}
0 &=& N e^{-\psi}   [  \tilde{\pi}^2 -\pi^a_i A_a^i  \pi^b_j A_b^j  ] 
+\partial_b   ( N A_a^i \tilde{E}^{a j} \tilde{E}^{bk} \epsilon_{ijk} ) 
\nonumber
\\
&&
 - \sqrt{ det( \tilde{\pi}^a_i ) }  < N e^{-\psi}   [  \tilde{\pi}^2 -(\pi^a_i A_a^j \pi^b_j A_b^i  ]    >
\nonumber \\
&&
-\frac{3}{2}  <\pi_\psi > e_{a}^i {\cal D}_b   ( N\tilde{E}^{a j} \tilde{E}^{bk} \epsilon_{ijk}   ) 
\label{Neq}
\end{eqnarray}
where $e_{a}^i$ is the dual frame field.  

\section{Concluding comments}

I close with brief comments on further work.  

\begin{itemize}

\item{}The results here are so far confined to Eucliean signature spacetimes.  The Plebanski action for Lorentzian signature  can be approached two ways: by complexifying the self-dual connection and then imposing reality conditions or by going to the full Lorentzian connection with a real Immirzi parameter.  There appears no barrier in principle to constructing the linking theory by going to the extended phase space in either case, but the details remain to be worked out.

\item{Loop representation for shape dynamics.}

One approach to quantum mechanics is to quantize the linking theory. This adds $\psi$ and $\pi_\psi$
to the observables of general relativity, along with an eighth first class constraint,
${\cal S}=0$.  
A starting point for loop quantization of the linking theory is the holonomy-flux algebra of the generalized holonomy variables discussed in section 2.3.  This is scale invariant.  The spectrum of areas is the original
spectrum multiplied by $e^{-\psi}$.  
This is a sense in which $e^{-\psi}$ functions as a dynamical Barbero-Immirzi parameter.
Apart from that, it is not clear what advantages such a quantization of the linking theory might have.

The alternative is to gauge fix and then quantize.  One can of course gauge fix back to general relativity,
and then quantize, but that is the same as quantizing general relativity to begin with.
Or we can try to gauge fix back to shape dynamics and then quantize.  This is a new direction, with
new challenges dude to the complexity of the Hamiltonian.

Once one fixes a gauge to descend from the linking theory, either areas nor volumes are Weyl invariant observables. 
There is however a tension between the local scale invariance of shape dynamics and the minimal areas and volumes in loop quantum gravity.  There are observables in LQG invariant under local scale transformations, these include the angles between edges  at nodes with valence of five or higher\cite{angles}.  However, Bianca Dittrich has pointed out that the Poisson bracket of two angle observables depends on scale dependent observables\cite{angles}.

If we restrict attention to four-valent graphs there are no diffeomorphism invariant angles at single nodes.  There are angle observables for large complex subgraphs using Penrose's original spin geometry theorem.   So if 
$\rho$ is a large subgraph of a graph $\Gamma$, connected to the rest by $N_\rho$ edges we can use Penrose's spin-geometry theorem to assign an angle
between any pairs of the outgoing edges. 

Tim Koslowski\cite{Tim-LSD} has suggested representing loop quantum shape dynamics by using unlabeled graphs.  So the space of states is the same as
$LQG$ without spin or intertwined labels. 

Once the states are understood, spin foam histories may be constructed from dual Pachner moves as in \cite{FM}.  As there are no labels there are only amplitudes labeled by the kind of move,
i.e,
\f
A_{1 \rightarrow 4},  \ \ \ A_{4 \rightarrow 1},  \ \ \ A_{2 \rightarrow 3}, \ \ \  A_{3 \rightarrow 2}
\ff
If we want time reversal invariance then there are two independent amplitudes
\f
A_{1 \rightarrow 4} = A_{4 \rightarrow 1},  \ \ \ A_{2 \rightarrow 3} =  A_{3 \rightarrow 2}
\ff
corresponding to the two independent constants, $G$ and $\Lambda$, i.e. the $1 \rightarrow 4$ moves generates expansion without shear and so is labeled  by
$\Lambda$ whereas the $2 \rightarrow 3$ and $3 \rightarrow 2$ moves are a mixture of expansion and shear and so are related to a combination of $G$ and $\Lambda$.

One goal is then to calculate correlation functions for perturbations of coarse grained angles to propagate.

\item{}A key open issue is whether the equations (\ref{weq},\ref{Neq})  have any solutions and, if so, whether they are unique.  It will also  be important to understand the relationship between these equations (\ref{weq},\ref{Neq}) and their counterparts in the $ADM$ formulation of shape dynamics.  The two sets of equations differ in that the latter have one more derivative than those here.  it may be that the reason is that one needs sot convert from first order form to second order form to convert connection dynamics into metric dynamics.  

\end{itemize}

\section*{ACKNOWLEDGEMENTS}

I am grateful to Tim Koslowski for discussions about his work on loop quantization of shape dynamics\cite{Tim-LSD}, as well as a careful reading of the paper, and to Sean Gryb for illuminating me on a subtle issue.  I would also  like to thank Julian Barbour, Bianca Dittrich, Henrique Gomes, Iwo Ita, Flavio Mercati and David Sloan for discussions and suggestions.  

This research was supported in part by Perimeter Institute for Theoretical Physics. Research at Perimeter Institute is supported by the Government of Canada through Industry Canada and by the Province of Ontario through the Ministry of Research and Innovation. This research was also partly supported by grants from NSERC, FQXi and the John Templeton Foundation.

\end{document}